\documentclass[aip,apl,amsmath,
amssymb,
 reprint,%
]{revtex4-1}
\usepackage{graphicx}
\usepackage{dcolumn}
\usepackage[latin9]{inputenc}
\usepackage{datetime} 
\usepackage{wasysym}
\usepackage{pifont} 
\usepackage{color} 
\usepackage{natbib}
\begin{document}

\title{Three-port beam splitter for slow neutrons using holographic nanoparticle-polymer composite diffraction gratings}

\author{J. Klepp} 
\email[]{juergen.klepp@univie.ac.at}
\homepage[]{http://fun.univie.ac.at} 
\affiliation{University of Vienna, 
Faculty of Physics, 1090 Wien, Austria}
\author{Y. Tomita}
\affiliation{University of Electro-Communications, 
Department of Engineering Science, 1-5-1 Chofugaoka, Chofu, Tokyo 182, Japan}
\author{C. Pruner}
\affiliation{University of Salzburg, 
Department of Materials Science and Physics, 5020 Salzburg, Austria}\author{J. Kohlbrecher} 
\affiliation{Laboratory for Neutron Scattering, Paul Scherrer Institut, 5232 Villigen PSI, Switzerland}
\author{M. Fally} 
\affiliation{University of Vienna, 
Faculty of Physics, 1090 Wien, Austria}
\date{\today} 


%

\begin{abstract}
Diffraction of slow neutrons by nanoparticle-polymer composite gratings has been observed. 
By carefully choosing grating parameters such as grating thickness and spacing, a three-port beam splitter operation for cold neutrons -- splitting the incident neutron intensity equally into the $\pm 1^{\mbox{\scriptsize{st}}}$ and $0^{\mbox{\scriptsize{th}}}$ diffraction orders -- was realized. As a possible application, a Zernike three-path interferometer is briefly discussed.
\end{abstract}

\maketitle

Since almost 40 years, neutron interferometers for thermal neutrons have played an important role in investigations of fundamental physics \cite{RauchWerner2000,HasegawaPRL2001,*
HasegawaNature2003,*HuberPRL2009,*
BartosikPRL2009,*pushinPRL2011}. 
The classic Mach-Zehnder geometry of the most successful perfect-crystal interferometers was also implemented using artificial structures, such as Ni gratings in reflection geometry combined with mirrors \cite{IoffePL1985}, or thin quartz glass phase gratings for very-cold neutron interferometry \cite{VanDerZouwNIMA2000}.
Multilayer-mirrors have been employed for cold neutron interferometry \cite{Funahashi-pra96}.
Gaining ground also as a tool in materials science, neutron interferometers are increasingly adopted for neutron phase-imaging and tomography \cite{zawiskyEPL2004,*PfeifferPRL2006,*pushinAPL2007,*
mankeNatCom2010}.
Many interesting effects of fundamental interest that -- due to their subtle nature -- are yet to be observed experimentally, can be investigated using neutron interferometers (see, for instance, Refs.\,\onlinecite{RauchWerner2000,wietfeldtPhysB2006,*
springerActaCryst2010,*pokotilovskiJETP2011}). In some cases, the rather small expected phase shifts can be enhanced by extending the interaction time of neutrons with the particular optical potential -- materials or magnetic- and electric-fields in one of the interferometer paths -- needed to probe a certain effect. Longer interaction times are achieved by enlarging the potential region and/or using slow neutrons with long de-Broglie wavelength. 
While the former is often difficult to realize for technical reasons or simply due to lack of space, the latter comes at the cost of neutron flux with today's neutron sources. Provided that efficient neutron-optical devices -- mirrors and beam splitters -- exist to avoid abolishing the advantage of slow-neutron interferometry by lack of intensity, cold and very-cold neutrons (0.5\,$\apprle\!\lambda\!\apprle$\,10nm) can offer a fair trade-off between neutron flux and interaction time. 
Mach-Zehnder interferometers for cold neutrons, based on holographic poly (methyl methacrylate) diffraction gratings, were tested in the past\cite{Schellhorn-phb97,*Pruner-nima06}. Only recently, it has been shown that holographic nanoparticle-polymer diffraction gratings are well suited for use as optical components in neutron interferometry. Such beam splitters\cite{FallyPRL2010,KleppPRA2011} and  free-standing film mirrors\cite{Klepp2012a} exhibit diffraction in the Bragg regime (two-wave regime) so that no intensity is wasted to undesired diffraction orders. 
In this Letter, a holographic SiO$_2$ nanoparticle-polymer neutron diffraction grating is presented that acts as a three-port beam splitter for cold neutrons at wavelength $\lambda=1.7$nm, i.e. the incident beam is coherently split into three beams of equal intensity.    

\begin{figure}
    \scalebox{0.36}
{\includegraphics {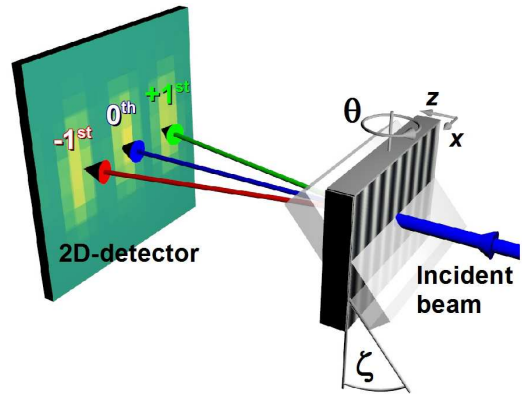}}
    \caption{Sketch of the experimental setup.}
    \label{fig1}
\end{figure}

Holographic gratings are produced by using a conventional holography setup. Coherent plane-wave signal and reference light-beams are superposed enclosing a certain crossing angle at the position of the recording material. Upon illumination with the resulting sinusoidal spatial light pattern, a sinusoidal neutron refractive-index modulation occurs in the recording material. It is proportional to the modulation of the coherent scattering length density $b_c\Delta\rho$.  
Such a holographic grating is
described by the periodically modulated refractive index 
$n(x)=\overline n+\Delta n \cos\left(2\pi x/\Lambda\right)$.
Here, $\Delta n=\lambda^2~ b_c\Delta\rho/(2\pi)$ and $\Lambda$ are the refractive-index modulation amplitude and the grating spacing, respectively. The quantities $b_c$ and $\Delta\rho$ are the coherent nuclear scattering length of the material and the number density modulation of the hologram, respectively.

Inorganic nanoparticles embedded in a photopolymer matrix (nanoparticle-polymer composites) have already been investigated \cite{SuzukiAPL2002,*SuzukiApplOpt2004,*
TomitaOptLett2005,*
SuzukiOptExp2006,*
SakhnoNanotech2007,*
NakamuraJOptA2009,*SakhnoJOptA2009} for light-optics applications.
A particular advantage of nanoparticle-polymer gratings -- expected to boost their impact on neutron optics -- is the possibility to tune the refractive-index modulation $\Delta n$ by choosing a suitable value of $b_c$, i.e. the species of nanoparticles. At the moment, a drawback is that due to detrimental holographic light-scattering in such thick composite films possessing large $\Delta n$ during recording, $\Delta n$ is reduced along the sample depth. As a result, grating thickness may be limited to the order of 100 microns, which is too small for reaching 50\% or 100\% reflectivity for cold and very-cold neutrons. Fortunately, optical elements are still feasible by exploiting the so-called Pendell\"{o}sung interference effect predicted by dynamical diffraction theory\cite{RauchWerner2000}: due to superposition of two neutron waves formed in the periodic potential of crystals and also holographic gratings, the output neutron intensity is swapped back and forth between diffracted and forward-diffracted (transmitted) beams, depending on the thickness\cite{Sippel-pl65} of the sample and the incident neutron wavelength\cite{Shull-prl68}. Thus, crystals or gratings can be tilted around an axis parallel to the grating vector\cite{SomenkovSolStComm1978}, thereby changing the effective thickness. This allows to control the diffraction efficiency (reflectivity) to some extent\cite{KleppPRA2011}.

The SiO$_2$ nanoparticles used for the present studies have an average core diameter of about 13\,nm. They were produced by liquid-phase synthesis and dissolved in a methyl isobutyl ketone solution. The SiO$_2$ sol was dispersed to methacrylate monomer. The nanoparticle concentration was 20 vol\%.
As radical photoinitiator, 1wt.\% titanocene (Irgacure784, Ciba) was added to enable the monomer to photo-polymerize at wavelengths shorter than 550 nm. The mixed syrup was cast on a 100-$\mu$m spacer-loaded glass plate and dried. Then the sample was covered with another glass plate to obtain a nanoparticle-monomer film.
Note that, prior to holographic recording, the photoinitiator, the monomer and the nanoparticles were homogeneously mixed in the sample material. 
Two expanded, mutually coherent and $s$-polarized laser beams at a wavelength of 532nm were superposed at a crossing angle of about 30$^\circ$ to create a spatially sinusoidal light-intensity pattern with a grating spacing of $\Lambda=1\,\mu$m in the sample. The beam diameter was about 1\,cm. 
The light pattern induced polymerization in the bright sample regions, a process that consumes monomers. As a consequence of the growing monomer-concentration gradient, 
nanoparticles counter-diffused from bright to dark regions\cite{TomitaOptLett2005}. This results in an approximately sinusoidal nanoparticle-concentration 
pattern. Homogeneous illumination after recording was made in order that the sample was fully polymerized and the nanoparticle density-modulation was fixed. It is found that similar samples stored under ambient conditions have not changed their light-diffraction properties for at least 7 years.

\begin{figure}
    \scalebox{0.51}
{\includegraphics {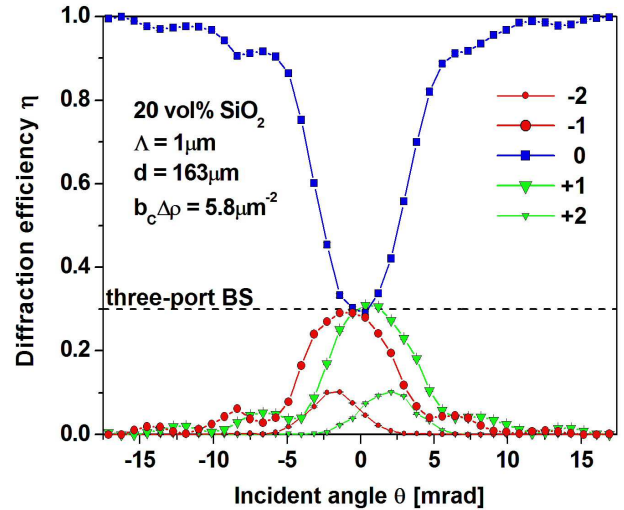}}
    \caption{Rocking curve measured at wavelength $\lambda=1.7$nm and grating tilt angle $\zeta=56$\textdegree. At $\theta=0$, the intensities of $\pm 1^{\mbox{\scriptsize{st}}}$ and $0^{\mbox{\scriptsize{th}}}$ diffraction orders are approximately equal and the grating acts as a three-port beam splitter.}
    \label{fig2}
\end{figure}

Neutron diffraction by such gratings was investigated at the instrument SANS I of the SINQ spallation source at the Paul Scherrer Institut (PSI) in Villigen, Switzerland.
The measurement principle is sketched in Fig.\,\ref{fig1}. 
The gratings were mounted in transmission geometry. Tilting the gratings to the angle $\zeta$ around an axis parallel to the grating vector -- in order to adjust the effective thickness -- the incident angle $\theta$ was varied to measure rocking curves in the vicinity of the Bragg angle $\theta_B$. 
The neutron wavelength distribution of the incident beam is typically 10\%.
The collimation-slit width and distance were chosen so that the beam divergence was about 1mrad. 
We defined the diffraction efficiency of the $i^{\mbox{\scriptsize{th}}}$ diffraction order as 
$\eta_{i}=I_{i}/I_{\mbox{\scriptsize{tot}}}$, 
where $I_i$ and $I_{\mbox{\scriptsize{tot}}}$ are the measured intensities of the $i^{\mbox{\scriptsize{th}}}$ diffraction order and the total intensity of transmitted and diffracted beams, respectively. For each incident angle $\theta$, the sum over all 2D-detector pixels in each separated diffraction spot (see Fig.\,\ref{fig1}) -- associated to one diffraction order -- was calculated and the obtained intensities (corrected for background) plugged into the above equation for $\eta_i$.
The resulting rocking curve is plotted in Fig.\,\ref{fig2}. 
Here, the incident neutron wavelength was $\lambda=1.7$nm and $\zeta=56$\textdegree. At this tilt angle, the effective thickness of the grating is about 163 microns. Note that in addition to the $\pm 1^{\mbox{\scriptsize{st}}}$ diffraction orders, the $\pm 2^{\mbox{\scriptsize{nd}}}$ diffraction orders are clearly visible. However, at $\theta=0$, the $\pm 2^{\mbox{\scriptsize{nd}}}$ diffraction orders decrease to $\eta_{\pm 2}\simeq 0.06$, while 
the $\pm 1^{\mbox{\scriptsize{st}}}$ and $0^{\mbox{\scriptsize{th}}}$ orders exhibit approximately the same values of about 0.3 . Thus, apart from negligible losses to the $\pm 2^{\mbox{\scriptsize{nd}}}$ diffraction orders, at normal incidence angle the incoming neutron wave is coherently split into three separated beams of equal intensity. The grating acts as a three-port beam splitter with overall efficiency $\eta_{-1}+\eta_{0}+\eta_{+1}$ of almost 0.9, which is close to the theoretical upper bound of about 0.93 \cite{goriOptComm1998}.
\begin{figure}
    \scalebox{0.44}
{\includegraphics {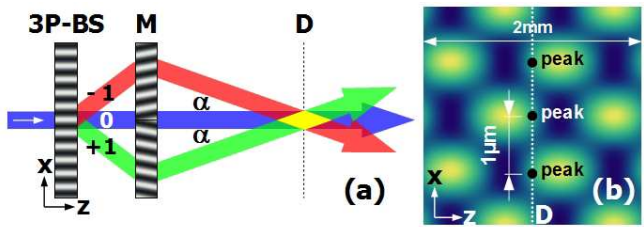}}
    \caption{(a) Sketch of a Zernike interferometer for very-cold neutrons with three-port beam splitter (3P-BS), and symmetrically slanted mirror grating (M). (b) Close-up of the expected interference pattern. Introducing a sample on the central beam, the intensity pattern is shifted along the $z$ direction, which results in a variation of peak intensities at the detection plane D, where the highest sensitivity is achieved.}
    \label{fig3}
\end{figure} 

A three-port beam splitter could be the central building-block of a Zernike three-path interferometer\cite{ZernikeJOptSocAm1950} for very-cold neutrons, as sketched in Fig.\,\ref{fig3}(a). 
Choosing suitable intensity ratios and phase relations between the three beams results in a neutron interference pattern with periodicity $\lambda/(2\sin\alpha)$ in $x$ direction, where $2\alpha$ is the angle enclosed by the $\pm 1^{\mbox{\scriptsize{st}}}$ order beams superposed at the detection plane [marked `D' in Fig.\,\ref{fig3}(a)]. An additional phase shift introduced on the central beam manifests as a shift of the interference pattern along the optical axis ($z$ direction) rather than a lateral shift of fringes. At the detection plane D, the introduced phase shift can be extracted from the resulting change of peak intensities of the observed pattern. To avoid aberration effects, recombination of the neutron beams at D can be achieved by using diffraction gratings as mirrors\cite{Klepp2012a} [marked `M' in Fig.\,\ref{fig3}(a)]. For the periodicity of the interference pattern in $x$ direction to be large enough for detection, two symmetrically slanted mirror-gratings can be recorded in one single nanoparticle-polymer composite sample in order to decrease the angle $2\alpha$. 
Suppose we require to maintain at least 1\,cm separation between the three beams (of 1\,mm width, say) at point M to conveniently insert a sample into the central beam. From geometrical considerations, it is found that for an incident neutron wavelength of $\lambda=8$nm and a three-port beam splitter with a grating spacing of $\Lambda=0.5\mu$m, the interference pattern formed along the $x$ direction at D can have a periodicity of $1\mu$m at a beam splitter-to-detector distance of about 3.75m. The pattern is spread 25cm along the optical axis and at most 1mm along the $x$ direction. A close-up of the expected intensity pattern is shown in Fig.\,\ref{fig3}(b). It can be observed by scanning the $x$ position of an absorption grating at D in front of a neutron detector. Neutron-absorption gratings of sufficiently small periodicity can already be produced by inclined sidewall evaporation of Gd on a Si grating\cite{GruenzweigRScIntr2008} or by holographic recording of CdSe nanoparticle-polymer gratings\cite{LiuAPL2009}, for instance. 
Note that -- for the given experimental parameters -- a phase shift as small as $\lambda/100$ introduced on the central beam is transferred to a spatial shift of the intensity pattern along the $z$ direction as large as $10\,\mu$m. Also, we do not expect overly stringent requirements for grating adjustment and beam collimation because nanoparticle-polymer composite gratings feature an angular selectivity -- approximately given by the width of the rocking curve $\Lambda/d$ -- that is reduced by about one order of magnitude as compared to the gratings for the interferometers described in Refs.\,\onlinecite{Schellhorn-phb97,*Pruner-nima06}. 

To conclude, we have demonstrated the implementation of a holographic SiO$_2$ nanoparticle-polymer composite grating as a three-port beam splitter for cold neutrons. 
As an example, we have briefly discussed its application in a Zernike three-path interferometer. Next to reaching high phase-sensitivity, such an instrument could also be used in a precision test of Born's rule\cite{SinhaScience2010,*deRaedtPRA2012} for matter-waves.

Financial support by the Austrian Science Fund (FWF): P-20265, and by the Ministry of Education,
Culture, Sports, Science and Technology of Japan (Grant
No. 23656045) is greatly acknowledged.


%

\end{document}